\documentclass[twocolumn,prb,amsmath,amssymb,showpacs]{revtex4}

\usepackage{graphicx}
\usepackage{dcolumn}
\usepackage{bm}

\begin{document}

\preprint{APS/V. Sih}

\title{Mechanical control of spin-orbit splitting in GaAs and InGaAs epilayers}

\author{V. Sih}
\author{H. Knotz}
\author{J. Stephens}
\author{V. R. Horowitz}
\author{A. C. Gossard}
\author{D. D. Awschalom}
\email{awsch@physics.ucsb.edu}
\affiliation{%
Center for Spintronics and Quantum Computation\\
University of California, Santa Barbara, CA 93106}

\date{\today}

\begin{abstract}
Time-resolved Kerr rotation spectroscopy as a function of
pump-probe distance, voltage and magnetic field is used to measure
the momentum-dependent spin splitting energies in GaAs and InGaAs
epilayers. The strain of the samples can be reproducibly
controlled in the cryostat using three- and four-point bending
applied with a mechanical vise. We find that the magnitude of the
spin splitting increases linearly with applied tension and
voltage. A strain-drift-diffusion model is used to relate the
magnitude of the measured spin-orbit splitting to the amount of
strain in the sample.

\end{abstract}

\pacs{71.70.Ej, 71.70.Fk, 72.25.Dc, 72.25.Rb}

\maketitle

Potential applications in spintronics~\cite{wolf01} and quantum
information processing~\cite{awsch02} rely upon an understanding
of the effect of electric fields and strain on electron spins.
Strain reduces the symmetry of a crystal, which introduces
momentum {\bf k}-linear terms to the Dresselhaus~\cite{dres} and
Bychkov-Rashba~\cite{rashba} spin splittings. These strain-induced
effective magnetic fields can be used to generate electron spin
polarization electrically~\cite{katoPRL} and coherently manipulate
spins using electric fields and in the absence of magnetic
fields~\cite{kato04}, but they also contribute to more efficient
spin relaxation~\cite{knotz}. In addition, recent steady-state
measurements~\cite{crooker05,beck} have shown that the spatial
period of strain-induced spin precession is independent of the
applied electric field, which demonstrates the robustness of
strain-induced spin precession for applications in functional
spin-based devices.

Here we employ mechanical three- and four-point bending to tune
the tensile strain of GaAs and InGaAs epilayers while performing
low-temperature time-resolved magneto-optical spectroscopy to
determine the magnitude of the strain-induced spin splitting. The
samples are contacted so that an in-plane electric field can be
applied to impart an average drift velocity to the
optically-excited electron spins. Kerr rotation measurements as a
function of magnetic field and pump-probe distance are performed
for different applied electric fields, and we observe that the
spin splitting increases with increasing drift velocity and
tensile strain. Unlike previous measurements that introduced
strain through heterostructure engineering and lattice-mismatched
growth~\cite{kato04}, these measurements are able to map out the
strain dependence in a single sample and without the complications
of strain relaxation. The vise geometry allows for repeatable
tensioning of samples and precise control over the strain level.

The samples are grown using molecular beam epitaxy on
semi-insulating (001) GaAs substrates. We examine both n-doped
GaAs and InGaAs epilayers. The GaAs samples are comprised of 100
nm undoped GaAs buffer layer, 400 nm Al$_{0.7}$Ga$_{0.3}$As, and a
500 nm Si-doped GaAs epilayer. Samples with carrier densities of
$2 \times 10^{16}$ cm$^{-3}$ and $4 \times 10^{16}$ cm$^{-3}$ were
measured, but since they exhibit qualitatively similar behavior,
we show only data for the $2 \times 10^{16}$ cm$^{-3}$ $n$-GaAs
sample in this paper. The InGaAs sample is composed of 300 nm of
growth-interrupted GaAs buffer layer, 500 nm of Si-doped
In$_{0.04}$Ga$_{0.96}$As with a carrier concentration of $3 \times
10^{16}$ cm$^{-3}$, and 100 nm of undoped GaAs. The lattice
constant of the InGaAs layer is matched to that of GaAs, as
confirmed using x-ray diffraction. Since InGaAs has a larger
natural lattice constant than GaAs, the InGaAs layer is
compressively strained in-plane as grown.

The samples are patterned into mesas using photolithography and a
chemical wet etch and then contacted using annealed Ni/AuGe
[Fig.~1(a)]. The channels have a width $w$ = 120 $\mu$m and length
$l$ = 310 $\mu$m between the contacts and are aligned such that an
electric field E = V/$l$ can be applied along either the
[1$\overline{1}$0] ($x$) or [110] ($y$) directions. The samples
are then mounted into either a three- or four-point mechanical
bending vise. While the four-point bending geometry produces
strain that is uniform between the central two
points~\cite{strain}, higher maximum values of strain can be
achieved in the three-point geometry before structural failure
occurs. In the case of the four-point bending, the mesas are
patterned far from the four contact points, minimizing local
strain variations. The sample is cooled to a temperature T = 60 K,
at which nuclear polarization is negligible~\cite{kikk00}. For the
field-dependent measurements, a magnetic field is applied along
$x$. A cross-section of the measurement geometry is shown in Fig.
1(b). The vise is tightened along the $z$ direction, introducing
tensile strain along $x$. The optical measurements are performed
along $z$ and measure the spin polarization along [001].

\begin{figure}
\includegraphics[width=0.44\textwidth]{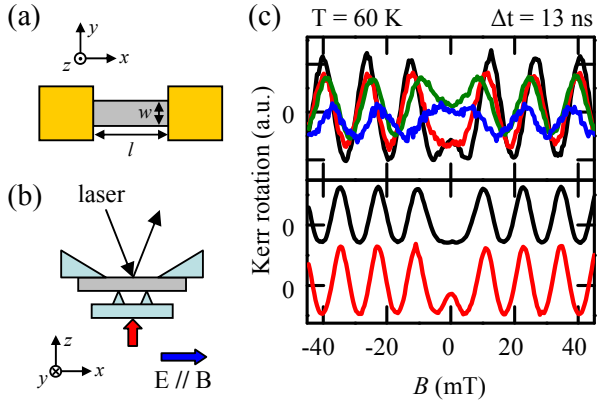}
\caption{\label{fig:epsart} (a) Patterned sample. Mesa is shown in
grey, and metal contacts are shown in gold. (b) Measurement
geometry showing orientation of sample, mechanical vise, electric
and magnetic fields and optical measurement. (c) (top) Kerr
rotation at temperature T = 60 K on $2 \times 10^{16}$ cm$^{-3}$
GaAs epilayer as a function of applied magnetic field for electric
field E = 32 V cm$^{-1}$ and pump-probe separation $d$ = 38 $\mu$m
at time delay $\Delta$t = 13 ns for strained states 2 (black), 3
(red), 4 (green) and 5 (blue). (bottom) Kerr rotation on InGaAs
epilayer for unstrained (black) and strained (red) states.}
\end{figure}

Time-resolved Kerr rotation spectroscopy~\cite{crooker97} is used
to monitor the electron spin dynamics in the samples. In this
technique, a Ti:Sapphire laser, tuned to the absorption edge of
the material that we wish to probe (wavelength $\lambda$ = 818 nm
for GaAs and $\lambda$ = 850 nm for In$_{0.04}$Ga$_{0.96}$As),
produces a train of $\sim$250 fs pulses at a repetition rate of 76
MHz, which are split into a pump beam (2 mW) and a probe beam (200
$\mu$W) with 30 $\mu$m diameters. The pump pulse is
circularly-polarized and excites a spin-polarized electron
population in the epilayer. A linearly-polarized probe pulse is
incident on the sample at time $\Delta$t later, which is
controlled using a mechanical delay line. The electron spin
polarization in the sample is measured by detecting the change in
the polarization axis, or Kerr rotation, of the reflected probe
beam. Time-resolved measurements at temperature T = 60 K and an
applied magnetic field B = 0.2 T show that the electron g-factor
is -0.43 for the GaAs epilayer and -0.48 for the InGaAs epilayer
and that the transverse spin coherence time is 40 ns for GaAs and
3 ns for the InGaAs sample. For the spatially-resolved
measurements, a stepper motor-driven mirror changes the spatial
separation $d$ between the pump and probe beams~\cite{kikk99}. The
applied electric field causes the spins to drift with an average
velocity $v_{d}$ and imparts a non-zero average momentum
$\langle${\bf k}$\rangle$ to the electron spin packet.

In order to determine the {\bf k}-dependent internal magnetic
field B$_{\mathrm{int}}$, we measure Kerr rotation as a function
of the applied magnetic field B$_{\mathrm{ext}}$ at $\Delta$t = 13
ns for various E and pump-probe distances. The presence of
B$_{\mathrm{int}}$ modifies the symmetric cosinusoidal
field-dependent signal~\cite{kato04,kikk99,kalevich}. When
B$_{\mathrm{int}}$ is along the same direction as
B$_{\mathrm{ext}}$, the signal becomes centered about
-B$_{\mathrm{int}}$, but if B$_{\mathrm{int}}$ and
B$_{\mathrm{ext}}$ are perpendicular, as is the case when the
electric field and B$_{\mathrm{ext}}$ are applied along the same
direction, the data can be fit to the equation:
\begin{eqnarray}
\mathrm{KR} = A
\mathrm{cos}(\text{g}\mu_{B}\sqrt{B_{\mathrm{ext}}^2 +
B_{\mathrm{int}}^2}\Delta\text{t}/\hbar)
\end{eqnarray}
where $A$ is the amplitude, g the effective g-factor of the
sample, $\mu_{B}$ the Bohr magneton, and $\hbar$ is Planck's
constant over 2$\pi$. For E = 32 V cm$^{-1}$ and $\Delta$t = 13
ns, the center of the electron spin packet is observed to be at
pump-probe separation $d$ = 38 $\mu$m. Figure 1(c) (top) shows
data for the GaAs sample for different amounts of tension applied
using the four-point bending vise. We measure Kerr rotation in the
channel for the unstrained case and for five increasing levels of
strain, which we label as strained states 1 - 5. The vise is
tightened by the same amount between each of these strained
states. Increasing the strain in the GaAs sample decreases the
signal amplitude and increases the spin precession frequency and
B$_{\mathrm{int}}$. The change in the amplitude is due to a
decrease in the spin lifetime, which we confirm using
time-resolved Kerr rotation. In contrast, measurements on InGaAs,
shown in Fig.~1(c) (bottom), show that the application of tensile
strain via the three-point bending vise leads to some relaxation
of the compressive strain due to lattice mismatch. The Kerr
rotation amplitude is increased, and B$_{\mathrm{int}}$ is
decreased. Measurements were also performed on the $2 \times
10^{16}$ cm$^{-3}$ GaAs sample for channels where E was applied
along $y$ and perpendicular to B$_{\mathrm{ext}}$ and the
direction of the tensile strain; in this geometry,
B$_{\mathrm{int}}$ is parallel to B$_{\mathrm{ext}}$, and the
values obtained for B$_{\mathrm{int}}$ were 22$\%$ smaller.

\begin{figure}
\includegraphics[width=0.44\textwidth]{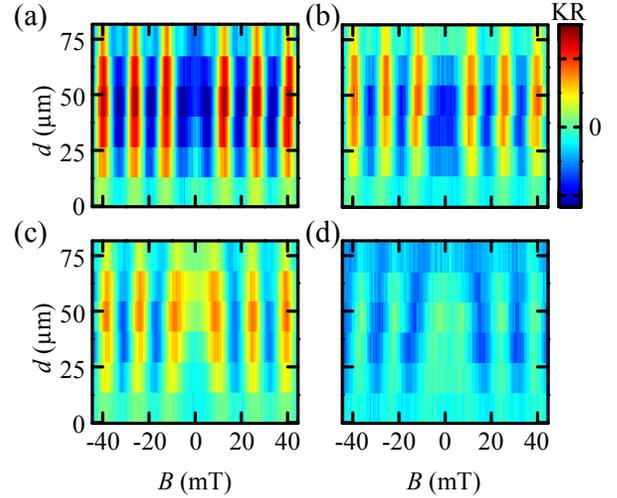}
\caption{\label{fig:epsart} Kerr rotation at temperature T = 60 K
and time delay $\Delta$t = 13 ns as a function of applied magnetic
field B$_{\mathrm{ext}}$ and pump-probe separation $d$ on $2
\times 10^{16}$ cm$^{-3}$ GaAs epilayer sample for electric field
E = 32 V cm$^{-1}$ for (a) strained states 2 (b) 3 (c) 4 and (d)
5. The color scale is the same for all four plots. }
\end{figure}

We explore the strain-dependent spin-orbit splitting in the GaAs
sample as a function of electric field and pump-probe distance. In
Figures 2(a)-(d), we show Kerr rotation as a function of applied
magnetic field and pump-probe separation for strained states 2, 3,
4, and 5 for E = 32 V cm$^{-1}$ and $\Delta$t = 13 ns. The color
scale is the same for all four plots. Again, we observe that the
spin precession period and amplitude decrease with strain and that
an increase in B$_{\mathrm{int}}$ lowers the central peak. In
these measurements, we also observe the effect of spin diffusion,
which is manifest in the spatial dependence of B$_{\mathrm{int}}$
and is due to the spread in drift velocities of the spin packet.

\begin{figure}
\includegraphics[width=0.44\textwidth]{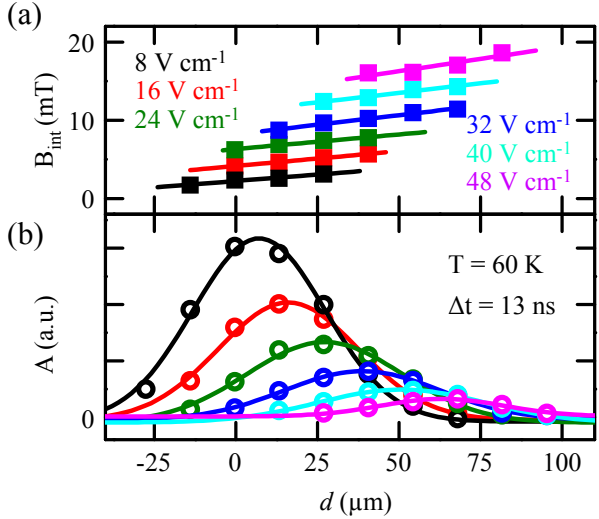}
\caption{\label{fig:epsart} (a) Effective magnetic field
B$_{\mathrm{int}}$ and (b) amplitude A as a function of pump-probe
separation $d$ on the $2 \times 10^{16}$ cm$^{-3}$ GaAs epilayer
sample in strained state 4 for electric field E = 8 (black), 16
(red), 24 (green), 32 (blue), 40 (cyan) and 48 V cm$^{-1}$
(magenta). Symbols are data, and lines are fits as described in
text. }
\end{figure}

In order to characterize the voltage-dependence of
B$_{\mathrm{int}}$, we fit the amplitude $A$ as a function of $d$
with a Gaussian function to determine the center position of the
spin packet $d_{c}$. We then use the results of a linear fit of
B$_{\mathrm{int}}$ to determine the value of B$_{\mathrm{int}}$ at
$d_{c}$ for each voltage. In Figure~3, we show the data (symbols)
and fits (lines) for $A$ and B$_{\mathrm{int}}$ for the GaAs
sample for strained state 4 as a function of $d$ and for various
values of E. From these fits, we obtain the spin-splitting energy
$\Delta_{0}$ = g $\mu_{B}$ B$_{\mathrm{int}}$ at the center of the
spin packet as a function of $v_{d}$, which is plotted in Fig.~4
for increasing amounts of strain in the $2 \times 10^{16}$
cm$^{-3}$ GaAs sample. As observed previously~\cite{kato04}, the
data can be fit to a line, where the slope $\beta$ =
$\Delta_{0}$/$v_{d}$ can be used to characterize the observed
effect. We plot $\beta$ for each of the strained states in the
inset of Fig.~4 and observe that $\beta$ increases for increasing
amounts of applied tension. In comparison, previous
measurements~\cite{kato04} on GaAs strained by the removal of the
underlying substrate showed that $\beta$ = 99 neV ns
$\mu$m$^{-1}$. Similar measurements of the InGaAs sample reveal
that $\beta$ = 72 neV ns $\mu$m$^{-1}$ when unstrained and $\beta$
= 40 neV ns $\mu$m$^{-1}$ when strained with the three-point
bending vise.

\begin{figure}
\includegraphics[width=0.44\textwidth]{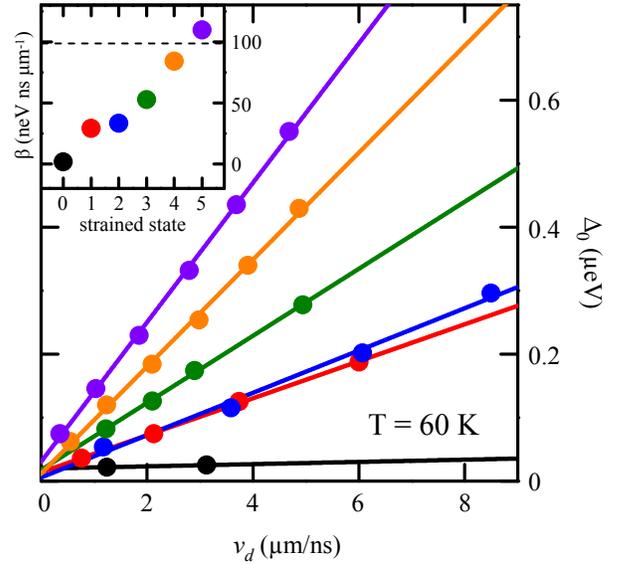}
\caption{\label{fig:epsart} Spin-splitting energy $\Delta_{0}$ as
a function of drift velocity $v_{d}$ for $2 \times 10^{16}$
cm$^{-3}$ GaAs epilayer sample when unstrained (black symbols) and
strained states 1 (red), 2 (blue), 3 (green), 4 (orange), and 5
(purple). Linear fits are also shown. (inset) $\beta$ [neV ns
$\mu$m$^{-1}$] for unstrained (0) and strained states 1 - 5. The
dashed line indicates $\beta$ from measurements on a GaAs membrane
in Ref.~[\onlinecite{kato04}].}
\end{figure}

Although we tightened the mechanical vise by the same amount
between each of the strained states, slip and play in the vise
make it difficult to determine the amount of bending and strain in
the sample by the mechanical displacement $\Delta$$z$. In order to
estimate the amount of strain for each of the strained states, we
solve a strain-drift-diffusion model~\cite{crooker05,hruska} to
determine the spatial spin precession period as a function of
$\varepsilon$. Although this model was developed for steady-state
measurements, as described in Ref.~[\onlinecite{crooker05}], using
the spatial spin precession period~\cite{kato05} SPP = 2 $\pi$
$\hbar$/ $\beta$, we can relate $\beta$ and $\varepsilon$. The
model parameters used are similar to those used in
Ref.~[\onlinecite{crooker05,hruska}] with the exception of the
spin diffusion constant $D$. A value of $D$ = 283 V/cm$^{2}$ was
obtained from the measurements in Fig.~3. The
strain-drift-diffusion equations are
\begin{eqnarray}
O_{1} \rho_{x} = O_{2} \rho_{z}
\end{eqnarray}
\begin{eqnarray}
O_{1} \rho_{y} = O_{3} \rho_{z}
\end{eqnarray}
\begin{eqnarray}
O_{4} \rho_{z} + O_{2} \rho_{x} + O_{3} \rho_{y} = -G_{z}
\end{eqnarray}
where the operators $O_{1}$-$O_{4}$ are defined as follows,
\begin{eqnarray}
O_{1} = D \nabla^{2} + \mu {\bf E} \cdot \nabla - (C_{S}D)^{2} -
\frac{1}{\tau_{s}}
\end{eqnarray}
\begin{eqnarray}
O_{2} = -C_{B_{y}} - C_{S}(2D\frac{\partial}{\partial x} + \mu
E_{x})
\end{eqnarray}
\begin{eqnarray}
O_{3} = C_{B_{x}} - C_{S}(2D\frac{\partial}{\partial y} + \mu
E_{y})
\end{eqnarray}
\begin{eqnarray}
O_{4} = D \nabla^{2} + \mu {\bf E} \cdot \nabla - 2 (C_{S}D)^{2} -
\frac{1}{\tau_{s}}
\end{eqnarray}

\begin{figure}
\includegraphics[width=0.44\textwidth]{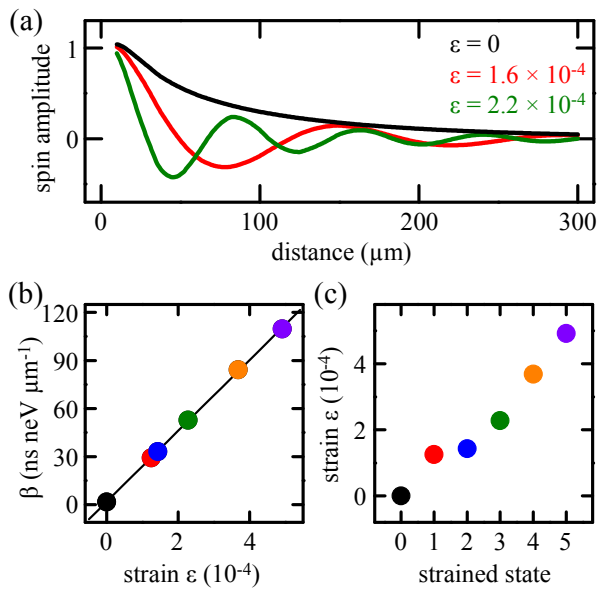}
\caption{\label{fig:epsart} (a) Spin polarization $\rho_{z}$ as a
function of distance solved using the strain-drift-diffusion model
for $\varepsilon$ = 0 (black), 1.6 (red) and 2.2 $\times$
10$^{-4}$ (green). (b) The linear relation between $\beta$ and
strain $\varepsilon$ from the strain-drift-diffusion model (line).
The symbols show $\beta$ obtained from measurements on the 2
$\times$ 10$^{16}$ cm$^{-3}$ GaAs epilayer sample. (c) Estimates
of strain $\varepsilon$ for the strained states from the measured
$\beta$. }
\end{figure}

Here $\rho_{i}$ is the $\{$x,y,z$\}$ component of electron spin
polarization, $G_{z}$ is a Gaussian source function with a FWHM of
30 $\mu$m, $D$ is the spin diffusion constant, $\mu$ is the
mobility, $\tau_{s}$ is the spin relaxation time, {\bf E} is the
applied electric field, {\bf B} is the applied magnetic field,
$C_{S}$ = $C_{3}$ m* $\varepsilon$/$\hbar^{2}$, $\hbar C_{B_{i}}$
= g $\mu$ B$_{i}$, m* is the electron effective mass, and
$\varepsilon$ is the strain. We assume a value for the spin-strain
coupling coefficient C$_{3}$ = 4.0 eV $\mathrm{\AA}$, as in
Ref.~[\onlinecite{hruska}]. The equations are solved over a 1
$\times$ 1 mm field using a finite element method. For the strain
calibration, the external applied magnetic field was set to zero,
and an electric field of 33 V cm$^{-1}$ was used. We obtain
solutions of the spin polarization as a function of position for
varying values of $\varepsilon$ between 0 and 0.001. Figure~5(a)
shows three line cuts of $\rho_{z}$ taken along the direction of
the electric field for different solutions with $\varepsilon$ = 0,
1.6, and 2.2 $\times$ 10$^{-4}$. We fit these line cuts to
determine the spatial spin precession frequency as a function of
$\varepsilon$. A linear fit of $\beta$ as a function of
$\varepsilon$ yields a slope of 22.27 $\pm$ 0.96 in units of
$\beta$ per 10$^{-4}$ unit strain. Figure~5(b) shows $\beta$ as a
function of $\varepsilon$. This relation allows us to assign
strain values to all six tension states using their measured
values of $\beta$ [Fig.~5(c)].

In summary, we have performed quantitative measurements of the
spin-splitting energy as a function of voltage and strain on
samples mounted in a mechanical vise. Using a
strain-drift-diffusion model, we are able to estimate the strain
in the devices and calibrate the observed spin-splitting. This
spin-splitting can be used to locally and coherently manipulate
electron spins and electrically drive spin
resonance~\cite{kato04,rashba03}.

We acknowledge support from AFOSR, DARPA/DMEA, NSF and ONR.
\\

\end{document}